\documentclass[conference]{IEEEtran}
\usepackage{amsmath,epsfig,bm,amssymb,amsthm}
\usepackage{psfrag}

\ifCLASSINFOpdf
\else
\fi
\usepackage{amsmath,epsfig,bm,amssymb,amsthm}

\begin{document}
%
\title{A NLLS Based Sub-Nyquist Rate Spectrum Sensing for Wideband Cognitive Radio}

\author{\IEEEauthorblockN{M. R. Avendi, K. Haghighi, A. Panahi and M. Viberg}
\IEEEauthorblockA{Department of Signal and Systems\\
Chalmers University of Technology\\
Gothenburg, Sweden\\
}
}


%


\maketitle

\begin{abstract}
\label{abs}
For systems and devices, such as cognitive radio and networks, that need to be aware of available frequency bands, spectrum sensing has an important role. A major challenge in this area is the requirement of a high sampling rate in the sensing of a wideband signal. In this paper a wideband spectrum sensing method is presented that utilizes a sub-Nyquist sampling scheme to bring substantial savings in terms of the sampling rate. The correlation matrix of a finite number of noisy samples is computed and used by a non-linear least square (NLLS) estimator to detect the occupied and vacant channels of the spectrum. We provide an expression for the detection threshold as a function of sampling parameters and noise power. Also, a sequential forward selection algorithm is presented to find the occupied channels with low complexity. The method can be applied to both correlated and uncorrelated wideband multichannel signals. A comparison with conventional energy detection using Nyquist-rate sampling shows that the proposed scheme can yield similar performance for SNR above 4 dB with a factor of 3 smaller sampling rate.
\end{abstract}


%
\IEEEpeerreviewmaketitle

\section{Introduction}
\label{se:intro}
Cognitive radio aims to enhance the utilization of the radio frequency (RF) spectrum through dynamic spectrum access. The motivation behind cognitive radio and networks is the scarcity of the available frequency spectrum and the increasing demand, caused by the emerging wireless applications for mobile users.  Due to the current static spectrum licensing scheme, spectrum holes or spectrum opportunities arise. Spectrum holes are defined as frequency bands which are allocated to licensed users, but in some locations and sometimes are not utilized by them. Therefore, they could be accessed by unlicensed users \cite{DSACRN}. As such, the first cognitive task is to develop wireless spectral detection and estimation techniques for sensing and identification of the available spectrum \cite{ZTCS}. 

Some well-known spectrum sensing techniques are energy detection (ED) \cite{edU}, matched filter and cyclostationary feature detection \cite{CRHykin} that have been proposed for narrow band sensing. In these techniques, the received signal is filtered with narrowband band-pass filters and sampled uniformly at the Nyquist rate. A decision then is made, based on the signal properties, to detect presence ($H_1$) or absence ($H_0$) of a primary user in the considered band \cite{CRN}.

Future cognitive radios should be capable of scanning a wideband of frequencies, in the order of few GHz \cite{CWBSSG}. In the wideband regime, the radio front-end can employ a bank of band-pass filters to select a frequency band and then exploit the existing techniques for each narrowband, but this method requires a large number of RF components \cite{ZTCS}.

Alternatively, in order to identify the locations of vacant frequency bands, the entire wideband is modelled as a train of consecutive frequency sub-bands \cite{CRN} and the total wideband is sampled classically or in a compressed way. After estimation of the spectrum from the obtained samples, the conventional spectrum sensing methods such as ED would be applied to detect the signal in each band. Classical sampling of a wideband signal needs high sampling rate ADCs, which have to operate at or above the Nyquist rate. Clearly, this is a major implementation challenge. Recent work based on compressive sampling has been proposed to overcome the problem of high sampling rates in \cite{CWBSSG}, \cite{ZTCS}. However, estimating the spectrum of a signal from its compressed samples is achieved by solving an optimization problem \cite{CWBSSG}, which is too complicated in comparison to our method. As we will see, in our method, estimation of the signal spectrum is skipped, and we directly detect the occupied channels from the sampled data in the time domain.

By using the fact that the wireless signals in open-spectrum networks are typically sparse in the frequency domain, in the previous work \cite{WSMM}, we proposed a wideband spectrum sensing method based on MUSIC algorithm that would bring substantial saving in terms of the sampling rate. However, we would like to improve the detection performance in smaller samples and lower SNRs.

In this work, the same sampling strategy as \cite{WSMM} is utilized to reduce the sampling rate far below the Nyquist rate. However, the estimator is replaced by a non-linear least square (NLLS) algorithm that would be applied to both correlated and uncorrelated multichannel signals. The frequency band of interest is divided into a finite number of spectral bands, and the occupied bands are estimated by considering the correlation matrix of the sampled data using a NLLS estimator. A theoretical expression is derived for the detection threshold based on the sampling parameters and noise power. Also, a sequential NLLS algorithm is utilized to reduce the complexity of implementation. 

The outline of the paper is as follows. The next section states the signal model and problem formulation. Section \ref{sec:model} introduces the spectrum sensing method and explains the functionality of each block in the model. In Section \ref{sec:sim}, comparison with the ED method and simulation results are presented and finally a conclusion is given in Section \ref{sec:con}.

\section{Problem Statement}
\label{sec:problem}
The received signal $x(t)$ is assumed to be an analog wideband sparse spectrum signal, bandlimited to $[0,B_{max}]$. The Nyquist rate for this signal is equal to $B_{max}$. However, the discussion is easily
adopted to real-valued signals supported on $-\frac{B_{max}}{2}, +\frac{B_{max}}{2}]$. Denote the Fourier transform of $x(t)$ by $X(f)$. Depending on the application, the entire frequency band is segmented into $L$ narrowband channels, each of them with bandwidth $B$, such that $B_{max}=L\times B$. The signal bands could be either correlated or uncorrelated with each other. The channels  are indexed from $0$ to $L-1$. Then, the frequency elements of the signal in each spectral band is represented by $X(f+rB)$, $f\in [0,B]$ where $0\le r\le L-1$ is the channel index. Those spectral bands which contain part of the signal spectrum are termed active channels, and the remaining bands are called vacant channels. Denote the number of such active channels by $N$. The indices of the $N$ active channels are collected into a vector 
\begin{equation}
\label{eq:chset}
\mathbf{b}=[b_1,b_2,\dots,b_N]
\end{equation}
which is referred to as the active channel set. 

\begin{figure}[t]
\psfrag {Spectrum} [][][.8]{PSD}
\psfrag {frequency[MHz]} [][][.8]{frequency[\mbox{MHz}]}
\centerline{\epsfig{figure=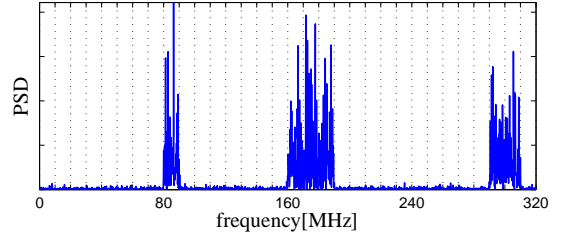},width=8.5cm}}
\caption{ Spectrum of a wideband signal received at the sensing radio with $L=32$ total bands and $N=6$ active channels. The active channel set is $\mathbf{b}=[8,16,17,18,29,30]$.}
\label{fig:sig}
\end{figure}

In the considered system, $N$ and $\mathbf{b}$ are unknown. 
However, we know the maximum percentage channel occupancy which is defined as 
\begin{equation}
\label{eq:occ}
\Omega_{max}=\frac{N_{max}}{L}
\end{equation}
where $N_{max}\ge N$ is the maximum possible number of occupied channels. The Landau's lower bound \cite{Landau} for this signal is equal to $\Omega_{max}\times B_{max}$. Figure ~\ref{fig:sig} depicts the spectrum of a multiband signal at the sensing radio, which contains $L=32$ channels, each with a bandwidth of $B=10$ MHz. The signal is present in $N=\nolinebreak[4]6$ channels, and the active channel set is $\mathbf{b}=[8,16,17,18,29,30]$.

The problem is, given $B_{max}$, $B$ and $\Omega_{max}$, to find the presence or absence of the signal in each spectral band or equivalently find the active channel set, $\mathbf{b}$, at a sub-Nyquist sample rate. 
\section{Wideband spectrum sensing model}
\label{sec:model}
\begin{figure*}[t]
\psfrag{xi[n]}{$x_i$}
\psfrag{x(t)}{$x(t)$}
\psfrag{Delay}{$z^{-c_i}$}
\psfrag{xi(n-c/L)}{$x_{d_i}$}
\psfrag{R=<x,x>}{$\frac{1}{M}\mathbf{x}_d\mathbf{x}_d^*$}
\psfrag{R}{$\mathbf{\hat{R}}$}
\psfrag{k}{${\mathbf{\hat{b}}}$}
\psfrag{Multicoset Sampler}{Multicoset Sampler}
\psfrag{Correlation matrix}{Sample Correlation matrix}
\psfrag{Subspace Methods}[.5]{$\qquad\qquad\qquad$NLLS estimator}
\psfrag{fs=a fmax}{$f_{avg}=\alpha B_{max}$}

\centerline{\epsfig{figure=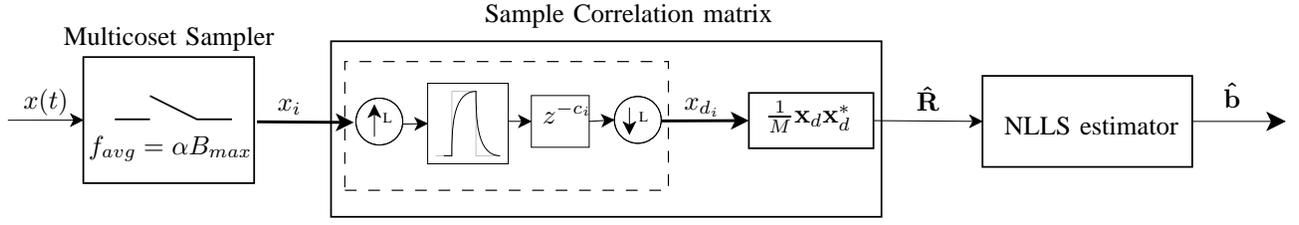},width=17cm}}
\caption{ Proposed wideband spectrum sensing model.}
\label{fig:model}
\end{figure*}

The proposed model for wideband spectrum sensing is illustrated in Figure ~\ref{fig:model}. The analog received signal at the sensing cognitive radio is sampled by the multicoset sampler at a sample rate lower than the Nyquist rate. The sampling reduction ratio is affected by the channel occupancy and multicoset sampling parameters. The outputs of the multicoset sampler are partially shifted using a multirate system, which contains the interpolation, delaying and downsampling stages. Next, the sample correlation matrix is computed from the finite number of obtained data. Finally, a NLLS estimator is used to discover the position of the active channels from the sample correlation matrix. In this section each block of the model is described in detail.

\subsection{Multicoset sampler}
\label{subsec:sampler}
The analog wideband signal $x(t)$ is sampled using a multicoset sampling scheme introduced and discussed in \cite{OptBres}, \cite{ThPrEldar}. The multicoset sampler provides $p$ data sequences for $i=1,...,p $, given by
\begin{equation}
 x_i(m)=x[(mL+c_i)/B_{max}],	m\in \mathbb{Z},
\end{equation}
where $c_i$, is a random number out of the set $\mathcal{L} =[0,1,...,L-1]$ \cite{OptBres}. 

The average sample rate of this scheme is $f_{avg} = \alpha B_{max}$, \cite{OptBres} where
$
\label{eq:alpha}
\alpha=(\frac{p}{L})
$
is termed the sub-Nyquist factor. According to Landau's lower bound \cite{Landau}, $\alpha$ is lower bounded to the maximum channel occupancy, $\alpha\ge \Omega_{max}$.

The number of data sequences, $p$, should be chosen greater than the maximum number of active channels $N_{max}$, to satisfy the Landau's lower bound and provide enough equations to find the unknown parameters.

\subsection{Sample correlation matrix}
\label{subsec:corr}
The main purpose of this section is to relate the problem of spectrum sensing with the problem of parameter estimation. Towards this goal, the correlation matrix of a special configuration of the sampled data is computed. In order to achieve this, the following configurations are applied on the sampled data. 

First, each $x_i(m)$ sequence is over-sampled by a factor $L$, such that 
$$
x_{u_i}[n]= \left\{\begin{array}{rcc}
x_i(\frac{n}{L}), &  n=mL, m\in \mathbb{Z}   \\ 0, & \mbox{otherwise}
\end{array}\right.
$$
and then it is filtered to obtain 
$
x_{h_i}[n]=x_{u_i}[n]* h[n],
$
where $h[n]$ is the interpolation filter with the frequency response of 
\begin{equation}
\label{eq:Hf}
H(f) = 
\begin{cases} 
1,  & f\in [0,B] \\
0, & \mbox{otherwise}. 
\end{cases}
\end{equation}
Next, the output filtered sequence is delayed with $c_i$ samples such that
\begin{equation}
\label{eq:xci}
x_{c_i}[n]=x_{h_i}[n-c_i].
\end{equation}

Let us define $\mathbf{y}(f)$ as the known vector of observations 
\begin{equation}
\label{eq:y}
\mathbf{y}(f)=
\begin{bmatrix}
X_{c_1}(f)\\
X_{c_2}(f)\\
\vdots \\
X_{c_p}(f)
\end{bmatrix}
\end{equation}
where $X_{c_i}(f)$ is the $DFT$ of the sequence $x_{c_i}[n]$. Also, $\mathbf{x}(f)$, the unknown vector of the signal spectrum parameters is defined as 
\begin{equation}
\label{eq:z}
\mathbf{x}(f)=
\begin{bmatrix}
{X}(f+b_1B)\\
{X}(f+b_2B)\\
\vdots \\
{X}(f+b_NB)\\
\end{bmatrix},
\quad f \in [0,B]
\end{equation}
where $X(f+b_iB),f\in [0,B]$, are the frequency elements of the signal in the active band indexed by $b_i$. 

After applying Fourier transform on both sides of (\ref{eq:xci}) and expressing the result in matrix form, the data model in the frequency domain is given by (Appendix A)
\begin{equation}
\label{eq:yaz}
\mathbf{y}(f)= \mathbf{A(b)} \mathbf{x}(f)+\mathbf{n}(f),\quad f \in [0,B]
\end{equation}
where $\mathbf{A(b)}\in \mathbb{C}^{p\times N}$ is the modulation matrix given by 
\begin{equation}
\label{eq:Ak}
\mathbf{A(b)}(i,k)= B \exp{\left(\frac{j2\pi c_i b_k}{L}\right)}
\end{equation}
and $\mathbf{n}(f)$ is the frequency representation of the noise. For simplicity, we assume that $\mathbf{n}(f)$ is a Gaussian complex noise with distribution of $\mathcal{N}(0,\sigma^2 \mathbf{I})$, which is also uncorrelated with the signal. 

The model in (\ref{eq:yaz}) is a classical signal model that relates the observation vector, $\mathbf{y}(f)$, with the unknown signal spectrum vector, $\mathbf{x}(f)$, via the modulation matrix $\mathbf{A(b)}$. Note that the unknown signal parameter $\mathbf{b}$ is also the active channel set that is desired in the problem of spectrum sensing. Therefore, with this configuration, the problem of wideband spectrum sensing is turned into the problem of finding the model parameter $\mathbf{b}$ with minimum length $N$, subject to the data model (\ref{eq:yaz}). This is a combined detection-estimation problem, where we want to estimate the number as well as the parameters of the signals. An approach based on the correlation matrix of the observations is employed here for solving this problem.

The correlation matrix of observation vector is defined as 
\begin{equation}
\label{eq:corr}
\mathbf{R}= E[\mathbf{y}(f)\mathbf{y}^*(f)]= \mathbf{A(b)} \mathbf{Q} \mathbf{A^*(b)}+ \sigma^2 \mathbf{I}
\end{equation}
where $( )^*$ denotes the Hermitian transpose, and 
\begin{equation}
\label{eq:Q}
\mathbf{Q}= E[\mathbf{x}(f) \mathbf{x}^*(f)]
\end{equation}
is the correlation matrix of the signal vector. 

Since the distribution of the signal is unknown, the real correlation matrix $\mathbf{R}$ cannot be achieved. Hence, we estimate $\mathbf{R}$ with an integration in the frequency domain as 
$$
\mathbf{\hat{R}}=\int\limits_{0}^{B} \mathbf{y}(f) \mathbf{y}^*(f) df
$$
However, $\mathbf{y}(f)$ is a vector of Fourier transform of $x_{c_i}$ samples, then from Parseval's identity, $\mathbf{\hat{R}}$ can be computed directly in the time domain from the  sequences $x_{c_i}[n]$ at the sample rate of $B_{max}$. Since each $x_{c_i}[n]$ sequence is the output of a narrowband filter, the reduced bandwidth output signal can be easily accommodated within a lower output sample rate. This means that the computations do not need to be performed at the high sample rate, $B_{max}$. Thus, the sequences are down-sampled by $L$, the reduced bandwidth factor, such that 
\begin{equation}
x_{d_i}(m)=x_{c_i}[mL]
\end{equation}

The total process of oversampling, filtering, delaying and downsampling from $x_i(m)$ to $x_{d_i}(m)$ is viewed as a fractional shifting of the sequence $x_i(n)$. The process also could be implemented in a very efficient way using polyphase filters \cite{polyphase}. Defining the snapshot vector $\mathbf{x}_d(m)$ as 
$$
\label{eq:xd}
\mathbf{x}_d(m)=
\begin{bmatrix}
x_{d_1}(m) \\
x_{d_2}(m) \\
\vdots \\
x_{d_p}(m)
\end{bmatrix},
$$ 
the $p\times p$ sample correlation matrix from $M$ samples of the partially shifted sequence is computed from the formula 
\begin{equation}
\label{eq:scorr}
\mathbf{\hat{R}}=\frac{1}{M} \sum_{m=1}^M \mathbf{x}_d(m) \mathbf{x}_d^*(m).
\end{equation}
Under suitable assumptions $\mathbf{\hat{R}} \rightarrow \mathbf{R}$ when $M\rightarrow \infty$.

\subsection{Least squares-based spectral estimation}
The sparse model in (\ref{eq:yaz}), to find a vector $\mathbf{b}$ with $N$ elements for some signals $\mathbf{x}(f)$, can be solved using a NLLS approach by minimizing the least square error (LSE) criterion as 
\begin{equation}
\label{eq:lse}
\providecommand{\norm}[1]{\lVert#1\rVert}
J(\mathbf{b},\mathbf{x}(f))= \int_0^B{\norm{\mathbf{y}(f)-\mathbf{A(b)}\mathbf{x}(f)}}^2_2df
\end{equation} 
where 
$
\providecommand{\norm}[1]{\lVert#1\rVert}
\norm{}_2
$
is the 2-norm vector. This is a separable least-squares problem, and for fixed (but unknown) $\mathbf{b}$, the solution with respect to the linear parameter $\mathbf{x}(f)$ is 
\begin{equation}
\label{eq:xhat}
\hat{\mathbf{x}}(f)= \mathbf{A}^{+} \mathbf{y}(f)
\end{equation}
where $\mathbf{A}^{+}=(\mathbf{A}^* \mathbf{A})^{-1} \mathbf{A}^*$ is the Pseudoinverse of $\mathbf{A(b)}$.

Substituting (\ref{eq:xhat}) into (\ref{eq:lse}) leads to \cite{FSSPV1} 
\begin{equation}
\label{eq:tr}
J(\mathbf{b})= tr\{(\mathbf{I}_p-\mathbf{A(b)}\mathbf{A}^{+})\mathbf{\hat{R}}\}
\end{equation}
where $\mathbf{I}_p$ denotes the identity matrix with dimension $p\times p$. 
Replacing $\mathbf{\hat{R}}$, by the true value (\ref{eq:corr}), shows that the minimum LSE is given by
\begin{equation*}
\begin{split}
J_{min}& =tr\{(\mathbf{I}_p-\mathbf{AA^{+}})\mathbf{(AQA^*+\sigma^2I)}\}\\
&= tr\{\mathbf{(\mathbf{I}_p-\mathbf{AA^{+}})\sigma^2I}\}\\
&= \sigma^2 (tr\{\mathbf{I}_p\}- tr\{\mathbf{AA^{+}}\})
\end{split}
\end{equation*}
It should be noted that $\mathbf{A}\in \mathbb{C}^{p\times N}$, $\mathbf{A}^*\in \mathbb{C}^{N\times p}$ and hence
\begin{equation*}
\begin{split}
tr\{\mathbf{AA^+}\}&= tr\{\mathbf{A^+A}\}\\
& =tr\{\mathbf{(A^*A)^{-1} A^*A} \} \\
&=tr\{\mathbf{I}_N\}=N
\end{split}
\end{equation*}
therefore,
\begin{equation}
J_{min}=\sigma^2 (p-N)
\label{eq:jmin}
\end{equation}
where $\sigma^2$ is the noise power. Therefore, with choosing $J_{min}$ as a detection threshold, the vector $\hat{\mathbf{b}}$ with the smallest length $\hat{N}$ that satisfies the condition 
\begin{equation}
\label{eq:th}
tr\{ (\mathbf{I}_p-\mathbf{\hat{A}A^+})\mathbf{\hat{R}}\}+\hat{N}\sigma^2 \le p \sigma^2
\end{equation}
where $\mathbf{\hat{A}=A(\hat{b})}$, is the solution of the spectrum sensing problem using NLLS method.

\begin{figure}[t]
\psfrag {J(b)} {\text{LSE}}
\psfrag {ls} [.7][]{$J(\mathbf{b}_{i+1})$}
\psfrag {i} {$i$}
\psfrag {jmin} {$J_{min}$}
\psfrag {th} {$(p-i)\sigma^2$}
\psfrag {Least Square Error} {}
\centerline{\epsfig{figure=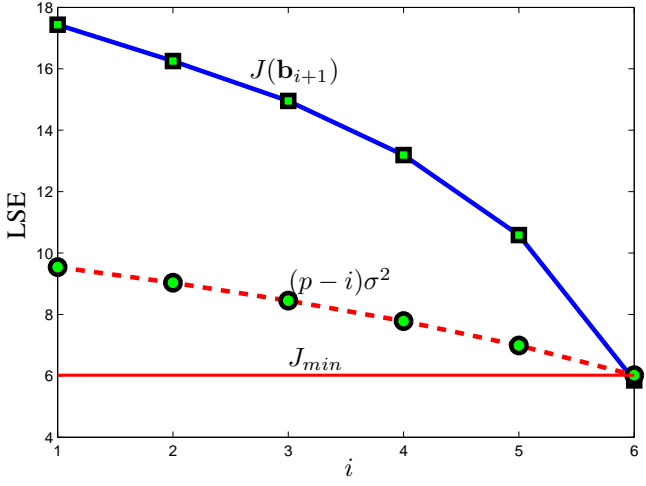},width=8.5cm}}
\caption{ The result of the sequential NLLS algorithm for a typical wideband system with $p=10, N=6, \sigma^2=1$.}
\label{fig:ls}
\end{figure}

The exhaustive search for finding $\hat{\mathbf{b}}$, at least needs $\sum_{i=1}^N \binom{L}{i}$ evaluations of (\ref{eq:th}), that is solvable only for small $N$ and $L$. A practical approach at a reasonable cost is to employ a sequential forward selection \cite{SAGE} where one channel of the spectral bands is selected at the time. The algorithm starts from the empty set, $\mathbf{b}_i=[\emptyset]$, and sequentially adds the channel index, $b^+$, that minimizes $J(\mathbf{b}_i \cup b)$. Meanwhile, the cell $b^+$ is augmented to the set; the value of least square criterion diminishes monotonically. The process is repeated until the criterion of (\ref{eq:th}) is satisfied at the perfect estimation point. Since the order of the model is unknown, the detection threshold in each step is determined by the expression $(p-i)\sigma^2$, where $i$ is the step index. The total number of evaluations in this way is less than $(L N)$. The process is summarized in the Algorithm 1.

\begin{table}[!t]
\renewcommand{\arraystretch}{2}
\label{tb:seqn}
\centering
\begin{tabular}{l}
\hline
\bf Algorithm 1: Sequential Forward NLLS\\
\hline
{\bf Input:} $\hat{\mathbf{R}}$ \\
{\bf Output:} $\hat{\mathbf{b}}, \hat{N}$ \\
1: Set $i=0, {\mathbf{b}}_i=[\emptyset]$ \\
2: Find $b^+= \arg \min \limits_{b \notin {\mathbf{b}}_i} J({\mathbf{b}}_i \cup b)$ \\
3: Update ${\mathbf{b}}_{i+1}={\mathbf{b}}_i \cup b^+, i=i+1$ \\
4: Go to step 2 if $J({\mathbf{b}}_{i+1})> (p-i)\sigma^2$ \\
5: {\bf return} $\hat{\mathbf{b}}={\mathbf{b}}_{i+1}, \hat{N}=i$ \\
\hline
\end{tabular}
\end{table}

Figure \ref{fig:ls} illustrates some important values of the sequential NLLS algorithm for a typical wideband system with $N=6$, $p=10$ (Figure \ref{fig:sig}). The exact detection threshold is shown by the horizontal line which is $J_{min}=(10-6)=6$ dB for the noise variance of $\sigma^2=1$. The LSE criterion starts at around $18$ dB and decreases monotonically, with adding any new cell to $\mathbf{b}_i$, until it surpasses the threshold level at $i=6$ which is the final estimation point. The estimated vectors in each step for the multichannel signal of Figure \ref{fig:sig} are listed below:

\begin{table}[!h]
\renewcommand{\arraystretch}{2}
\label{tb:bhat}
\begin{tabular}{l}
$\mathbf{b}_1= [18]$ \\
$\mathbf{b}_2= [18,29]$ \\
$\mathbf{b}_3= [8,18,29]$ \\
$\mathbf{b}_4= [8,17,18,29]$ \\
$\mathbf{b}_5= [8,17,18,29,30]$ \\
$\mathbf{b_6}= [8, 16, 17, 18, 29, 30]$ 
\end{tabular}
\end{table}

Where $\hat{N}=6$ and $\hat{\mathbf{b}}=\mathbf{b}_{6}$ are the outputs.

\section{Comparison and Simulation Results}
\label{sec:sim}
In this section, we illustrate the performance of the proposed scheme by comparing to an ED technique and also the previous method \cite{WSMM} using Monte Carlo simulations.

The received signal at the cognitive radio sensing is generated
from the model
$$
x[n]= \sum_{i=1}^{N} (r_i[n]*h[n]) \exp(j2\pi f_i n /B_{max})+ w[n]
$$
where $*$ shows a convolution between $r_i[n]\sim \mathcal{N}(0,\sigma_i^2)$ and $h[n]$, the low pass filter defined in (\ref{eq:Hf}). The output of the convolution is placed at the carrier frequency $f_i$, and corrupted by $w[n] \sim\mathcal{N}(0,1)$, the additive white Gaussian noise. 

The wideband of interest is in the range of $[0,320]$\nolinebreak[4] MHz, containing $32$ channels of equal bandwidth of $B=10$MHz. The signal variance is chosen such that the received SNR of all active channels are equal. Figure ~\ref{fig:sig} depicts the spectrum of the signal model with $N=6$ active bands located at different unknown carriers. Given $B_{max}=320$MHz, $\Omega_{max}= 0.25$ and $B=10$MHz, it is desired to find the positions of occupied and vacant channels at a sub-Nyquist sampling rate. 

A multicoset sampler with parameters $L=32, p=10$ is used to sample the signal at the average sample rate of  $f_{avg}=\nolinebreak[4]100$ MHz, which is $\alpha \approx 0.3$ of the Nyquist rate. Ten $c_i$ numbers are selected randomly out of the set $\mathcal{L}$. The correlation matrix is computed from $M=64$ configured samples and then the sequential NLLS algorithm is applied to estimate the vector $\mathbf{\hat{b}}$.

\begin{figure}[t]
\psfrag {x(t)} [][.8]{$x(t)$}
\psfrag {xi[n]} [] [.8]{$x(nT)$}
\psfrag {Multicoset Sampler} [][][.8] {Uniform Sampler}
\psfrag {fs=a fmax} [][][.8]{$f_s=B_{max}$}
\psfrag {Correlation matrix} {Filter Bank}
\psfrag {xh1} {$$}
\psfrag {xh2} {$$}
\psfrag {R=<x,x>} [][][.8]{$\frac{1}{M} \sum |.|^2$}
\psfrag {th} [][][.8]{$\gtrless_0^1 \eta $}
\psfrag {H0} {$H_0$}
\psfrag {H1} {$H_1$}
\centerline{\epsfig{figure=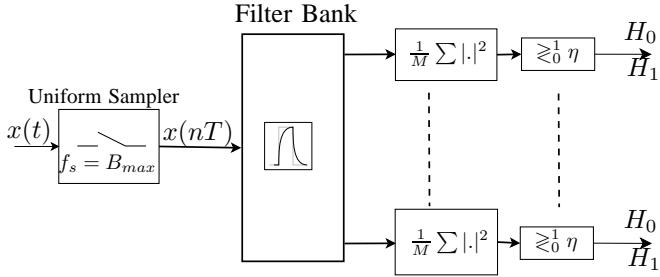},width=8.5cm}}
\caption{ Block diagram of the conventional ED method.}
\label{fig:ed2}
\end{figure}

The block diagram of the conventional ED model is illustrated in Figure \ref{fig:ed2}. As it is seen, the input signal is sampled conventionally at the Nyquist rate and then is filtered using a filter bank consisting of $L=32$ non-overlapping filters. The centre frequency of each filter corresponds to one spectral band in the signal spectrum. Then, the output signal energy is estimated from $M$ samples and is compared with a threshold, $\eta$, to decide if a signal is present $(H_1)$ or not $(H_0)$. The threshold is determined to meet the given false alarm probability $P_{\text{fa}}$ as \cite{ED1}
$$
\eta= \sigma^2 \left(1+\frac{Q^{-1}(P_{\text{fa}})}{\sqrt{M/2}}\right)
$$ 
where $\sigma^2$ is the noise power and $Q^{-1}$ is the inverse Q-function. Here, it is computed for $P_{\text{fa}}=0.01$, $\sigma^2=1$ and $M=64$ samples.

The detection performance is evaluated by computing the probability of detecting the signal occupancy as 
\begin{equation}
\label{eq:pdk}
P_{\mbox{d}}=\frac{1}{N} \sum_{i=1}^N \Pr(b_i\in \mathbf{\hat{b}}|b_i\in \mathbf{b})
\end{equation}  
and the false alarm probability as 
\begin{equation}
\label{eq:pfk}
P_{\mbox{f}}=\frac{1}{L-N} \sum_{i=1}^{L-N} \Pr(b_i^c\in \mathbf{\hat{b}}|b_i^c\in \mathbf{b}^c)
\end{equation}
where $\mathbf{b}^c= \mathcal{L} -\mathbf{b}$ is the complement set of $\mathcal{L}$. 

\begin{figure}[t]
\psfrag {detection}[][][.8]{$P_{\text{d}}$}
\psfrag {SNR} [][][.8]{$\text{SNR},[dB]$}
\centerline{\epsfig{figure=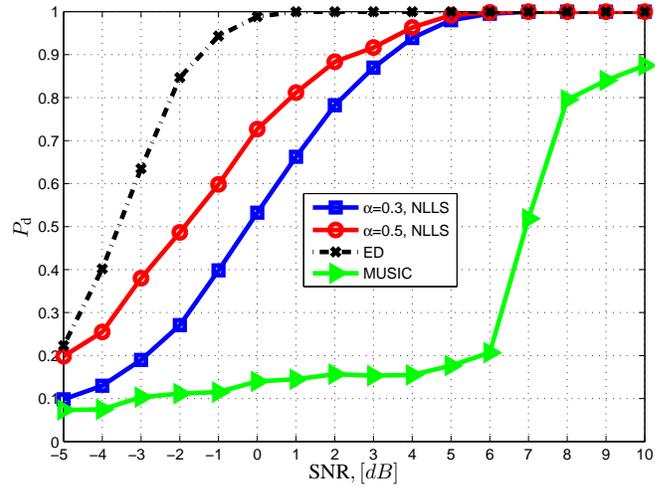},width=8.5cm}}
\caption{ Detecting probability of the proposed model and the ED method versus SNR for the simulated wideband system.}
\label{fig:pls}
\end{figure}

\begin{figure}[t]
\psfrag {detection}[][][.8]{$P_{\text{f}}$}
\psfrag {SNR} [][][.8]{$\text{SNR},[dB]$}
\centerline{\epsfig{figure=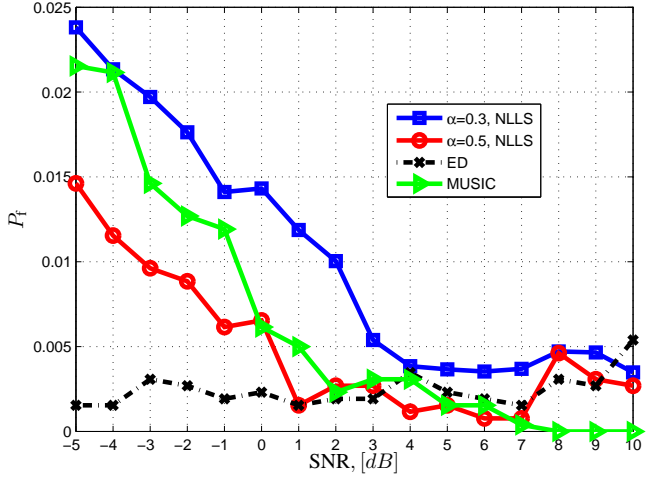},width=8.5cm}}
\caption{ False alarm probability of the proposed model and the ED method versus SNR for the simulated wideband system.}
\label{fig:pfls}
\end{figure}

In a comparison test, using 1000 Monte Carlo simulations, the same signal at various values of SNR is applied on all methods. The sensing period of three techniques is also the same. We compute the empirically observed $P_{\text{d}}$ and $P_{\text{f}}$. The result is shown in Figure \ref{fig:pls} and Figure \ref{fig:pfls}. We use the result of the ED method as a bench mark. It is seen that the performance of the NLLS method is superior to that of the MUSIC method, and it improves monotonically with increasing SNR. At $\alpha=0.3$ of the Nyquist rate, after SNR=4dB, the NLLS method is able to detect the occupied channels with high probability. At $\alpha=0.5$ of the Nyquist rate, this would improve especially for lower SNR and get closer to the ED result. Moreover, the corresponding $P_{\mbox{f}}$ values show almost a similar performance for SNR above 4dB. It should be noted that the empirical $P_{\mbox{f}}$ is averaged over the total vacant channels, hence it is lower than the desired value, ($P_{\mbox{fa}}=0.01$), of a single channel. No need to say, in addition to a high sample rate, the requirement of a large number of filters with different centre frequencies make serious implementation challenges in the ED method. This is replaced by a simple low pass filter in our method.

\begin{figure}[t]
\psfrag {Pd}[][][.8]{$P_{\text{d}}$}
\psfrag {M} [][][.8]{$M, \text{samples}$}
\centerline{\epsfig{figure=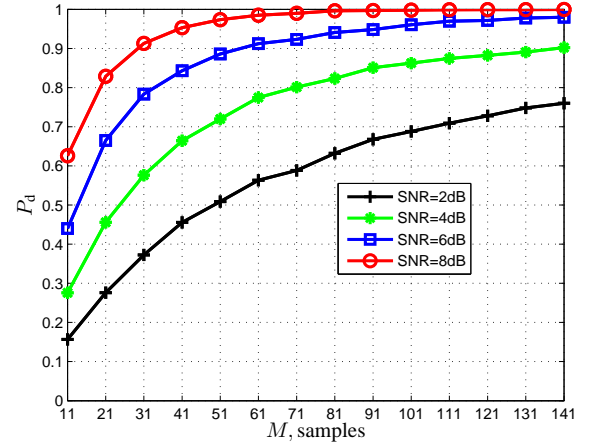},width=8.5cm}}
\caption{ Detecting performance of the proposed model versus the sensing period for the simulated wideband system.}
\label{fig:pdfM}
\end{figure}

In another experiment, the simulation is repeated for different number of samples to see the effect of sensing period on the detection performance of proposed method. The $P_{\text{d}}$ for various values of $M$ and SNRs is computed again and plotted in Figure~\ref{fig:pdfM}. As it is seen, the detection performance is improved dramatically with increasing $M$. However, we would see a wall for large values of $M$.

\section{Conclusion}
\label{sec:con}
We proposed a method of wideband spectrum sensing to mitigate the limitations of high sampling rate. The proposed technique utilizes a multicoset sampling scheme that can use arbitrarily low sampling rates close to the channel occupancy. With low spectrum utilization assumption, this would bring substantial savings in terms of the sampling rate. We modelled the problem as an overcomplete linear model with a sparse solution. We solved this sparse regression problem by a stepwise forward selection method. For model order selection we used the theoretical assumptions for multiple snapshots. The simulation results show that the proposed method has a promising detection performance, comparable to that of a conventional ED technique at high SNR. In our method, for a typical wideband system with $\Omega_{max}=0.25$ at SNR=5 dB and $\alpha\approx 0.3$ of the Nyquist rate, $P_{\mbox{d}}=0.98$ and $P_{\mbox{f}}= 0.004$ are achieved.

\section*{appendix A \\proof of (\ref{eq:yaz})}
\label{sec:appa}
Assume $B_{max}=1$ then $B=1/L$, the sampling sequences are given by
$$
x_i(m)= x(mL+c_i)
$$
Taking the DFT of the above results in
$$
X_i(f)=\frac{1}{L} \sum\limits_{r=0}^{L-1} X(\frac{f}{L}+\frac{r}{L}) \exp(\frac{j2\pi c_if}{L}) \exp(\frac{j2\pi c_i r}{L}) 
$$
where $X_i(f)$ is the DFT of $x_i(m)$ sequence.
Oversampling each sequence by $L$, such that
$$
x_{u_i}[n]= \left\{\begin{array}{rcc}
x_i(\frac{n}{L}), &  n=mL, m\in \mathbb{Z}   \\ 0, & \mbox{otherwise}
\end{array}\right.
$$
in the frequency domain we have
\begin{equation*}
\begin{split}
X_{u_i}(f)&=X_i(Lf)\\
&=B \sum\limits_{r=0}^{L-1} X(f+rB) \exp(j 2\pi c_if) \exp(\frac{j2\pi c_i r}{L})
\end{split}
\end{equation*}
where $X_{u_i}(f)$ is the DFT of the oversampled sequence $x_{u_i}$.
Filtering $X_{u_i}(f)$ with $H(f)$, limits the output signal frequency range such that
$$
X_{h_i}(f)=X_{u_i}(f), \quad f\in [0,B]
$$
and then delaying each sequence with $c_i$ samples gives
\begin{equation*}
\begin{split}
X_{c_i}(f)&= X_{h_i}(f) \exp(-j2\pi c_i f)\\
&= B \sum\limits_{r=0}^{L-1} X(f+rB)\exp(\frac{j2\pi c_i r}{L}),\quad f\in [0,B]
\end{split}
\end{equation*}
where $X_{c_i}(f)$ is the DFT of the delayed sequence $x_{c_i}$.

Assume the input signal $x(t)$ is perturbed by Gaussian noise $w(t)$, with Fourier transform $W(f)$. In the above equation, the right hand side expression is repeated for noise such that
\begin{equation*}
\begin{split}
X_{c_i}(f)&= B \sum\limits_{r=0}^{L-1} X(f+rB)\exp(\frac{j2\pi c_i r}{L}) \\
&+ B \sum\limits_{r=0}^{L-1} W(f+rB)\exp(\frac{j2\pi c_i r}{L}),\quad f\in [0,B] 
\end{split}
\end{equation*}
where the term $X(f+rB), \quad 0\le r\le L-1$, denotes the frequency elements of the signal in each channel which is zero for the vacant channels. So the equation is rewritten as
\begin{equation*}
\begin{split}
X_{c_i}(f)&=B \sum_{r\in \mathbf{b}} X(f+rB) \exp(\frac{j2\pi c_ir}{L})\\
&+ B \sum\limits_{r=0}^{L-1} W(f+rB)\exp(\frac{j2\pi c_i r}{L}),\quad f\in [0,B] 
\end{split}
\end{equation*}
Expressing the results in the matrix form for $X_{c_i}(f), i=1,\dots p$, we have 
\begin{equation*}
\mathbf{y}(f)= \mathbf{A(b)x}(f)+ \mathbf{n}(f),\quad f\in [0,B] 
\end{equation*}
where $\mathbf{y}(f),\mathbf{x}(f) \text{ and } \mathbf{A(b)}$ are defined earlier and $\mathbf{n}(f)$ is equivalent to the noise part.

\bibliographystyle{IEEEbib}
\bibliography{strings}

\end{document}